\documentclass[12pt]{article}
\usepackage{amsmath}
\usepackage{mathrsfs}
\usepackage{graphicx}
\usepackage{ amssymb }
\usepackage{multirow}
\usepackage[margin=2cm]{geometry} 
\usepackage{slashed}
\usepackage{caption}
\newcommand{\overbar}[1]{\mkern1.5mu\overline{\mkern-1.5mu#1\mkern-1.5mu}\mkern 1.5mu}

\def\un#1{\relax\ifmmode\@@underline#1\else
        $\@@underline{\hbox{#1}}$\relax\fi}

\let\du=\du                     

\def\c{\chi}

\def\f{\phi}

\def\l{\lambda}
\def\m{\mu}
\def\n{\nu}

\def\p{\pi}

\def\x{\xi}

\def\L{\Lambda}
\def\O{\Omega}

\def\ve{\varepsilon}

\def\bo{{\raise-.3ex\hbox{\large$\Box$}}}               
\def\pa{\partial}                                       
\def\TH{{\raise.2ex\hbox{$\displaystyle \bigodot$}\mskip-4.7mu \llap H \;}}
\def\face{{\raise.2ex\hbox{$\displaystyle \bigodot$}\mskip-2.2mu \llap {$\ddot
        \smile$}}}                                      
\def\dg{\sp\dagger}                                     

\def\sp#1{{}^{#1}}                              
\def\Bar#1{\overline{#1}}                       
\def\abs#1{\left| #1\right|}                    
\def\leftrightarrowfill{$\mathsurround=0pt \mathord\leftarrow \mkern-6mu
        \cleaders\hbox{$\mkern-2mu \mathord- \mkern-2mu$}\hfill
        \mkern-6mu \mathord\rightarrow$}
\def\dvec#1{\vbox{\ialign{##\crcr
        \leftrightarrowfill\crcr\noalign{\kern-1pt\nointerlineskip}
        $\hfil\displaystyle{#1}\hfil$\crcr}}}           

\newskip\humongous \humongous=0pt plus 1000pt minus 1000pt

\newif\ifdtup

 \newcommand{\be}{\begin{equation}}
\newcommand{\ee}{\end{equation}}
\newcommand{\nbe}{\begin{equation*}}
\newcommand{\nee}{\end{equation*}}

\newcommand{\lb}{\label}

\def\fracmm#1#2{{{#1}\over{#2}}}
  
\def\low#1{{\raise -3pt\hbox{${\hskip 0.75pt}\!_{#1}$}}}

\begin{document}

\thispagestyle{empty}

{\hbox to\hsize{
\vbox{\noindent November 2019 \hfill IPMU19-0158 }}
\noindent  \hfill }

\noindent
\vskip2.0cm
\begin{center}

{\Large\bf On the equivalence between Starobinsky and Higgs 
\vglue.1in inflationary models in gravity and supergravity
}

\vglue.3in

Sergei V. Ketov~${}^{a,b,c}$
\vglue.1in

${}^a$~Department of Physics, Tokyo Metropolitan University, \\
1-1 Minami-ohsawa, Hachioji-shi, Tokyo 192-0397, Japan \\
${}^b$~Research School of High-Energy Physics, Tomsk Polytechnic University,\\
2a Lenin Avenue, Tomsk 634028, Russia \\
${}^c$~Kavli Institute for the Physics and Mathematics of the Universe (WPI),
\\The University of Tokyo Institutes for Advanced Study, Kashiwa 277-8583, Japan  \\
\vglue.1in
ketov@tmu.ac.jp
\end{center}

\vglue.3in

\begin{center}
{\Large\bf Abstract}
\end{center}
\vglue.1in

\noindent  Starobinsky inflation and Higgs inflation in gravity, and their equivalence based on the common inflationary potential are extended to supergravity in the proper framework, where the Starobinsky and Higgs descriptions of inflation arise in two different gauges of the same supergravity model. 

\newpage

\section{Introduction}

The Starobinsky inflation \cite{Starobinsky:1980te} based on the modified $(R+R^2)$ gravity and the so-called Higgs inflation \cite{Bezrukov:2007ep} based on the non-minimal coupling of a Higgs field to gravity  are apparently different but, nevertheless, lead to the same predictions about inflation. Their equivalence is known in the literature and is based on the fact that both models have the same inflationary potential. 

In this paper we review both inflationary models in the gravitational theory by emphasizing the underlying assumptions leading to their equivalence in describing inflation, and then generalize this equivalence to supergravity theory by proposing the supergravity model where both (Starobinsky and Higgs) descriptions of inflation arise in two different gauges. 

The paper is organized as follows. In Sec.~2 we review the Starobinsky model of inflation in modified gravity and scalar-tensor gravity and comment on its essential features. In Sec.~3 we review the Higgs inflationary model with a nonminimal coupling to gravity and comment on its essential features. The known equivalence between the
two inflationary models in gravity theory is confirmed in Sec.~4. Several different ways of embedding the Starobinsky and Higgs inflationary models to supergravity are considered in Sec.~5, where we select the most economical supergravity framework for their realization.  In Sec.~6 we establish the equivalence between the supergravity-based Starobinsky and Higgs inflationary models. Sec.~7 is our Conclusion. 

\section{Review of Starobinsky inflation in modified gravity}

The purpose of this Section is to review the Starobinsky (1980) model of inflation and argue about its privileged position amongst all inflationary models. The Starobinsky model is defined by the action \cite{Starobinsky:1980te}
 \be \label{star}
S_{\rm Star.} = \fracmm{M^2_{\rm Pl}}{2}\int \mathrm{d}^4x\sqrt{-g} \left( R +\fracmm{1}{6m^2}R^2\right)~,
\ee
where we have introduced the reduced Planck mass $M_{\rm Pl}=1/\sqrt{8\p G_{\rm N}}\approx
2.4\times 10^{18}$ GeV, and the mass parameter $m$. We use the
spacetime signature $(-,+,+,+,)$. 

In the low curvature regime, the $R^2$ term can be ignored and the action (\ref{star}) reduces to the standard Einstein-Hilbert action. In the high curvature regime (relevant for inflation), the $R$ term can be ignored and the action  (\ref{star}) 
reduces to the no-scale $R^2$ gravity with the dimensionless coupling constant in front of the action. The $R^2$ coupling
must have a positive coupling constant to avoid a ghost.

The $(R+R^2)$ gravity model (\ref{star}) can be considered as a representative of the modified $F(R)$ gravity theories defined by 
 \be \label{fg}
S_F = \fracmm{M^2_{\rm Pl}}{2}\int \mathrm{d}^4x\sqrt{-g} \, F(R)
\ee
with a function $F(R)$ of the scalar curvature $R$.

At first sight, the model (\ref{star}) is rather {\it ad hoc}, being just one of many possible choices of the function $F(R)$.
However, a closer theoretical inspection and the current observational data strongly favor  Eq.~(\ref{star}) as the basic model of inflation. In order to demonstrate that, first, we recall the well known fact that the modified gravity theories can be reformulated as the scalar-tensor gravity theories \cite{mf}, see also Ref.~\cite{Aldabergenov:2018qhs} for a recent account and more details of computation.

The $F(R)$ gravity action (\ref{fg}) is classically equivalent to 
\begin{equation} \lb{eq}
 S[g_{\m\n},\chi] = \fracmm{M^2_{\rm Pl}}{2} \int d^{4}x \sqrt{- g}~ \left[ F'(\chi) (R - \chi) + F(\chi) \right]
\end{equation}
with the real scalar field $\chi$, provided that $F''\neq 0$ that we always assume, and the primes denote the derivatives with respect to the argument. The $\chi$-field equation of motion implies $\chi=R$ that brings back the action  (\ref{fg}).
Otherwise, the (positive) factor $F'$ in front of the $R$ in (\ref{eq}) can be eliminated by a Weyl transformation of metric $g_{\m\n}$, which transforms the action (\ref{eq}) into an action of the dynamical scalar field  $\chi$ minimally coupled to Einstein gravity and having the scalar potential 
\be \lb{spot}
 V =      \left(\fracmm{M^2_{\rm Pl}}{2}\right)       \fracmm{\chi F'(\chi) - F(\chi)}{F'(\chi)^{2}}~~.
\ee

The kinetic term of $\chi$ becomes canonically normalized after the field redefinition $\chi(\varphi)$ as
\begin{equation} \lb{fred}
  F'(\chi) =  \exp \left(  \sqrt{\frac{2}{3}} \varphi/M_{\rm Pl} \right)~,\quad
  \varphi =  \fracmm{\sqrt{3}M_{\rm Pl}}{\sqrt{2}}\ln F'(\c) ~~,
\end{equation}
in terms of the canonical inflaton field $\varphi$. It results in the scalar-tensor gravity (or quintessence) acton
\be \lb{quint}
S_{\rm quintessence}[g_{\m\n},\varphi]  = \fracmm{M^2_{\rm Pl}}{2}\int \mathrm{d}^4x\sqrt{-g} R 
 - \int \mathrm{d}^4x \sqrt{-g} \left[ \frac{1}{2}g^{\m\n}\pa_{\m}\varphi\pa_{\n}\varphi
 + V(\varphi)\right]~.
\ee

The classical and quantum stability conditions of $F(R)$ gravity theory (that we always assume) are given by 
\cite{Ketov:2012yz}
\begin{equation} \lb{stab}
 F'(R) > 0 \quad {\rm and} \quad F''(R) > 0~,
\end{equation}
while they are obviously satisfied in the case of the Starobinsky model (\ref{star}) for $R>0$. Actually, the first condition
in Eq.~(\ref{stab}) means that graviton is not a ghost, whereas the second condition means that inflaton is not a tachyon.

For our purposes in this paper, it is the {\it inverse} transformation that is more illuminating. It takes the form 
\cite{starp,Ketov:2014kya,Ketov:2014jta}
\begin{align} 
  & R = \left[  \fracmm{\sqrt{6}}{M_{\rm Pl}}
    \frac{d V}{d \varphi} + \fracmm{4V}{M^2_{\rm Pl}} \right] \exp \left(  \sqrt{\frac{2}{3}} 
  \varphi/M_{\rm Pl} \right),  \lb{inv1} \\
  & F= \left[  \fracmm{\sqrt{6}}{M_{\rm Pl}}
 \frac{d V}{d \varphi} + \fracmm{2V}{M^2_{\rm Pl}} \right] \exp \left(  2 \sqrt{\frac{2}{3}} \varphi/M_{\rm Pl}\right)~,
 \lb{inv2}
\end{align}
and defines the function $F(R)$ in the parametric form for a (given) inflaton scalar potential $V(\varphi)$. 

The key physical requirement to the inflaton potential is its {\it flatness} enabling slow roll of inflaton during sufficient duration of inflation (see below their precise definitions). It exactly corresponds to the smallness of the first term versus the second one in the square brackets of Eqs.~(\ref{inv1}) and (\ref{inv2}). When ignoring the first term, i.e. assuming $V\approx const.$, it immediately gives rise to the $F$-function as the $R^2$ term driving inflation in the Starobinsky model (\ref{star}).

As regards the $(R+R^2)$ gravity of Eq.~(\ref{star}), the exact inflaton potential is given by
\begin{equation} \label{starp}
V(\varphi) = \fracmm{3}{4} M^2_{\rm Pl}m^2\left[ 1- \exp\left(-\sqrt{\frac{2}{3}}\varphi/M_{\rm Pl}\right)\right]^2
\end{equation}
and has a {\it plateau} of positive height (related to the inflationary energy density), which gives rise to slow roll of inflaton indeed. 
 
A duration of inflation is measured by the e-foldings number
\be \lb{efold}
N_e\approx  \fracmm{1}{M^2_{\rm Pl}} \int_{\varphi_{\rm end}}^{\varphi_{*}} \fracmm{V}{V'}d\varphi~~,
\ee
where $\varphi_{*}$ is the inflaton value at the reference scale (horizon crossing), and $\varphi_{\rm end}$ is the
inflaton value at the end of inflation when one of the slow roll parameters,
\be \lb{slowp}
\ve_V(\varphi) = \fracmm{M^2_{\rm Pl}}{2}\left( \fracmm{V'}{V}\right)^2 \quad {\rm and} \quad 
\eta_V(\varphi) = M^2_{\rm Pl} \left| \fracmm{V''}{V}\right|~~,
\ee
is no longer small (close to 1).

The amplitude of scalar perturbations at horizon crossing is given by \cite{Ellis:2015pla}
\be \lb{amp}
A = \fracmm{V_*^3}{12\p^2 M^6_{\rm Pl}({V_*}')^2}=\fracmm{3m^2}{8\p^2M^2_{\rm Pl}}\sinh^4\left(
\fracmm{\varphi_*}{\sqrt{6}M_{\rm Pl}}\right)~~.
\ee

As regards phenomenological applications, the Starobinsky model  (\ref{star}) is the excellent model of cosmological inflation, in very good agreement with the Planck data (2018). The Planck satellite mission measurements of the Cosmic Microwave Background (CMB) radiation \cite{Planck18}
give the scalar perturbations tilt as $n_s\approx 1+2\eta_V -6\ve_V\approx 0.9649\pm 0.0042$ (with 68\% CL) and restrict the  tensor-to-scalar ratio as $r\approx 16\ve_V < 0.064$ (with 95\% CL). The Starobinsky inflation yields $r\approx 12/N_e^2\approx 0.004$ and $n_s\approx 1- 2/N_e$, where $N_e$ is the e-foldings number between 50 and 60, with the best fit at $N_e\approx 55$ \cite{Mukhanov:1981xt,Kaneda:2010ut}.

The Starobinsky model (\ref{star}) is geometrical (based on gravitational interactions only), while its (mass) parameter $m$ is fixed by the observed CMB amplitude (COBE, WMAP) as 
\be \lb{starm}
m\approx 3 \cdot10^{13}~{\rm GeV} \quad {\rm or}\quad \fracmm{m}{M_{\rm Pl}}\approx 1.3\cdot 10^{-5}~.
\ee
A numerical analysis of (\ref{efold}) with the potential (\ref{starp}) yields \cite{Ellis:2015pla}
\be\lb{sandf}
\sqrt{\fracmm{2}{3}} \varphi_*/M_{\rm Pl} \approx \ln\left( \fracmm{4}{3}N_e\right) \approx 5.5\quad {\rm and}
\quad \sqrt{\fracmm{2}{3}} \varphi_{\rm end}/M_{\rm Pl} \approx \ln\left[ \fracmm{2}{11}(4+3\sqrt{3})\right]\approx 0.5~~,
\ee
where we have used $N_e\approx 55$.

The scalar potential (\ref{starp}) is obviously {\it nonrenormalizable}. Expanding it in powers of  $\varphi/M_{\rm Pl}$ clearly shows that the nonrenormalizable (marginal) terms beyond the fourth power of $\varphi$ are suppressed by the powers of $M_{\rm Pl}$ so that the {\it UV-cutoff} of the Starobinsky model is given by $\L_{\rm Star.UV}= M_{\rm Pl}$, in agreement
with \cite{Hertzberg:2010dc}. The classical equivalence of $F(R)$ gravity and scalar-tensor gravity was extended to their
(on-shell) {\it quantum} equivalence in the 1-loop approximation in Ref.~\cite{Ruf:2017xon}.

\section{Review of Higgs inflation with non-minimal coupling to gravity}

The basic idea of Higgs inflation is to identify inflaton field with Higgs field. Strictly speaking, it does not have to be the Higgs particle of the Standard Model but just the Higgs potential for inflaton. It is phenomenologically possible when inflaton is non-minimally coupled to gravity \cite{Bezrukov:2007ep}. 

The Lagrangian (in Jordan frame) reads (we take $M_{\rm Pl}=1$ fo simplicity)
\be \lb{bs}
{\cal L}_J = \sqrt{-g} \left[ \frac{1}{2} (1+\x\phi^2)R - \frac{1}{2} g^{\m\n}\pa_{\m}\phi\pa_{\n}\phi - V_H(\phi) \right]~, 
\ee
where the scalar potential takes the Higgs form:
\be \lb{poth}
V_H(\phi) =\fracmm{\l}{4}\left( \phi^2 -v^2\right)^2~,
\ee
with the coupling constants $v$ and $\l$, while the non-minimal coupling to gravity (in front of the $R$) is measured by the new coupling constant $\x>0$.

A transfer from Jordan frame to Einstein frame is achieved via a Weyl transformation,
\be \lb{fje}
 g_J^{\m\n}= g_E^{\m\n}(1+\x\phi^2)~,
\ee
which results in a non-canonical kinetic term of the scalar $\phi$ and the rescaled scalar potential.

A canonical scalar kinetic term is obtained via a field redefinition  $\varphi=\varphi(\phi)$ according to
\be \lb{cano}
 \fracmm{d\varphi}{d\phi} =\fracmm{\sqrt{1+\x(1+6\x)\phi^2}}{1+\x\phi^2}
\ee
that gives rise to the standard Lagrangian 
\be \lb{bs2}
{\cal L}_E = \sqrt{-g} \left[ \frac{1}{2} R - \frac{1}{2} g^{\m\n}\pa_{\m}\phi\pa_{\n}\phi - V(\varphi) \right] 
\ee
with the scalar potential 
\be \lb{pote}
V(\varphi) =\fracmm{V_H(\phi(\varphi))}{\left[1+\x\phi^2(\varphi)\right]^2}~.
\ee

In the {\it large} field approximation, $\f^2\gg v^2$, a solution to Eq.~(\ref{cano}) takes the form
\be \lb{solc}
\varphi \approx \sqrt{\frac{3}{2}} \ln \left(1+\x\phi^2\right)
\ee
that yields  the scalar potential
\be \lb{potbs}
V(\varphi) = \fracmm{\l}{4\x^2}\left(1-e^{-\sqrt{2/3}\varphi}\right)^2
\ee
which  {\it coincides} with the Starobinsky inflationary potential (\ref{starp}).

The CMB observations require $\x/\sqrt{\l}\approx 5\cdot 10^4$ with the inflaton mass 
$m=\sqrt{\frac{\l}{3}}\x^{-1}\approx 10^{-5}$. It is usually assumed that the Higgs coupling constant $\l$
is of the order one, which implies that $\x$ is of the order $10^4$.

Unlike the renormalizable scalar potential (\ref{poth}) in Jordan frame, an expansion of the scalar potential (\ref{pote}) in powers of $\varphi$ in Einstein frame gives rise to the nonrenormalizable (marginal) 
terms multiplied by powers of $\x$. Therefore, after restoring the Planck scale  $M_{\rm Pl}$ by dimensional considerations, we conclude that the {\it UV-cutoff} of the Higgs inflationary model is given by  $\L_{\rm Higgs.UV}= M_{\rm Pl}/\x$, in
agreement with \cite{Burgess:2009ea}. The UV-cutoff $\L_{\rm Higgs.UV}$  is considerably lower the  UV-cutoff of the Starobinsky model $\L_{\rm Star.UV}= M_{\rm Pl}$ by the large factor $\x$, which implies that the Higgs inflation is more sensitive to quantum corrections than the Starobinsky inflation.

\section{Review of the equivalence between the Starobinsky and Higgs inflationary models} 

The coincidence of the inflationary potentials in the Starobinsky and Higgs inflationary models implies the same physical predictions for large-field inflation. This appears to be rather surprising, given the obvious differences between the definitions of those models. A simple explanation is possible after taking into account the slow roll condition  \cite{Ketov:2012jt,Kehagias:2013mya}. It is important to clearly identify the assumptions  leading to the equivalence because they define the range of its applicability.

First, the Higgs field $H$ of the Standard Model is a doublet i.e., it is a complex scalar field. However, one can choose the 
unitary gauge $H=\phi/\sqrt{2}$ in the Higgs Lagrangian ($M_{\rm Pl}=1$)
\be \lb{bsh}
{\cal L}_H = \sqrt{-g} \left[ \frac{1}{2}  R +\x H^{\dagger}HR - g^{\m\n}\pa_{\m}H^{\dagger}\pa_{\n}H - 
 \l\left(H^{\dg}H-\frac{1}{2} v^2\right)^2 \right]~, 
\ee
where the Higgs field can be represented by the real field $\phi$.

In the {\it large} field approximation one can ignore $v$ in the Higgs scalar potential, while during {\it slow roll} inflation one can also  ignore the scalar kinetic term against the scalar potential. Taken together, these two assumptions greatly simplify the above Lagrangian to
 \be \lb{bsha}
{\cal L}_H\approx \sqrt{-g} \left[ \frac{1}{2} (1+\x\phi^2)R - \fracmm{\l}{4}\phi^4 \right]~,
\ee
where the field $\phi$ becomes an auxiliary field. Varying the action with respect to $\phi$ yields $\x\phi R =\l \phi^3$ or
just 
\be \lb{sola}
\phi^2 = \fracmm{\x}{3}R~.
\ee 
Substituting this result back into the Lagrangian gives the Starobinsky model:
\be \lb{r2a}
{\cal L}_H\approx \sqrt{-g}\left( \fracmm{1}{2} R + \fracmm{\x^2}{4\l}R^2\right)
\ee

The established correspondence is known in the literature as the {\it asymptotic duality} between the Higgs and Starobinsky models of inflation. There is no correspondence in the small field approximation. Reheating is also known to be different. For instance, the reheating temperature $T_{\rm Higgs}\approx 10^{13}$ GeV, whereas $T_{\rm Star.}\approx 10^{9}$ \cite{Ketov:2012yz}. The difference in the UV-cutoff scales was already mentioned above.

We conclude this Section by an observation that the common inflationary potential (\ref{starp}) can be rewritten to the pure {\it mass term} after the field redefinition 
\be \lb{sred}
\fracmm{3}{4} m^2\left[ 1- \exp\left(-\sqrt{\frac{2}{3}}\varphi\right)\right]^2 = \fracmm{1}{2} m^2f^2
\ee
that gives rise to the non-canonical kinetic term of the field $\sqrt{2/3}f=1- \exp{\left(-\sqrt{\frac{2}{3}}\varphi\right)}$ as
\be \lb{skint}
-\fracmm{1}{2} \fracmm{(\partial f)^2}{(1-\sqrt{2/3}f)^2}~~.
\ee
The plateau of the canonical scalar potential (\ref{starp}) corresponds to {\it the pole} in the kinetic term of the inflaton field $f$. Therefore, the inflationary features are related to the critical phenomena in the $f$-theory (or the singularity theory), so that their universal behavior is no longer surprising  \cite{Ketov:1995yd}.

\section{Starobinsky and Higgs inflation in supergravity}

There are many ways of embedding the existing inflationary models to supergravity, see e.g., Refs.~\cite{Ketov:2012yz,Yamaguchi:2011kg} for a review, while it is not our purpose to review the literature here. In this Section, we briefly mention the problems related to the known approaches and choose the supergravity framework that is most suitable for our needs in this paper, namely the {\it minimal} realizations of Starobinsky inflation and Higgs inflation in the four-dimensional $N=1$ supergravity. 

The standard approach to inflation in supergravity is based on (local) supersymmetrization of a scalar-tensor gravity
with the quintessence action (\ref{quint}), while it assigns inflaton scalar to a {\it chiral} supermultiplet  with the  maximal spin 1/2 \cite{Yamaguchi:2011kg}. It requires {\it a complexification} of the inflaton scalar field by supersymmetry. Actually, two chiral superfields are generically needed:  the one including inflaton and another one including spin-1/2 {\it goldstino} related to spontaneous SUSY breaking caused by inflation. Therefore, the standard approach actually deals with a multi-field inflation, because two chiral superfields already have four physical scalars. It is possible to identify the inflaton chiral superfield with the goldstino chiral superfield  \cite{Ketov:2014qha,Ketov:2014hya} but even in that case inflaton is mixed with its pseudo-scalar superpartner in two-field inflation.
 
Slow-roll inflation is realized by {\it engineering} the scalar potential $V$ in terms 
of a K\"ahler potential $K$  and a superpotential $W$ of the chiral superfields as $(M_{\rm Pl}=1)$
\begin{equation}\label{eng}
V_F = e^K \left( \abs{DW}^2 - 3\abs{W}^2 \right) \quad  {\rm with} \quad  DW=W' +K'W~~,
\end{equation}
where the primes stand for the derivatives with respect to the chiral superfields (we use the same notation for the superfields and their leading field components).

There are several problems with the standard approach in supergravity: (i) the $\eta$-problem related to the $e^K$ factor in the scalar potential, in order to achieve flatness of the inflationary potential or, equivalently, get a small slow-roll parameter $\eta$, (ii) justification of a choice of the inflationary (single-field) trajectory in the (scalar) field space, and (iii) potential instabilities of the inflationary trajectory against mixing of inflaton with other scalars that may easily spoil single-field inflation, in order to achieve a desired number of e-foldings. Though all these problems can be solved, the predictive power of the standard approach is not high because the solutions usually require even more physical degrees of freedom and more 'ad hoc' interactions, with the infinitely many choices of $K$ and $W$.

It is also possible to supersymmetrize the modified $F(R)$ gravity theories in supergravity by using the (curved) superspace techniques \cite{Ketov:2010qz,Ketov:2013dfa}. A generic action is given by
\begin{equation}
\label{A}
S=\int d^{4}x d^{4}\theta E^{-1}N(\mathcal{R},\bar{\mathcal{R}})+\left[\int d^{4}x d^{2}\Theta 2\mathcal{E}F(\mathcal{R})+h.c    \right]
\end{equation}
and is governed by two potentials, $N$ and $F$, of the chiral superfield $\mathcal{R}$ having the scalar curvature $R$  as its field component at $\Theta^2$. The action (\ref{A}) is a sum of the $D$-type and $F$-type manifestly supersymmetric terms, respectively, though the (covariantly) chiral superfield $\mathcal{R}$ is not independent but obeys the
certain constraints in superspace \cite{Wess:1992cp}. As a result, the "auxiliary" fields of the supergravity multiplet become
physical degrees of freedom in the higher-derivative action (\ref{A}).

The action (\ref{A}) can be transformed into the {\it standard} matter-coupled supergravity action with two chiral matter superfields having the K\"ahler potential and the superpotential in terms of the input potentials $N$ and $F$
\cite{Cecotti:1987sa,Gates:2009hu}. Those (Einstein) supergravities with chiral matter have the same problems mentioned above (though in the very different parameterization), and are generically not free of ghosts and tachyons, which makes their selection to be a very difficult task \cite{Ketov:2013dfa}.

It is possible to extend the non-minimal Higgs-gravity coupling to supergravity too \cite{Ketov:2012jt}. 
The most general action of a chiral Higgs superfield $H$ coupled to supergravity (in Jordan frame, without higher derivatives) is given by
\begin{equation}
\label{su}
S_H=-3\int d^{4}x d^{4}\theta E^{-1}\exp\left[-\frac{1}{3}K(H,\Bar{H})\right]
+\left[\int d^{4}x d^{2}\Theta 2\mathcal{E}W(H)+h.c.   \right]
\end{equation}

The desired Higgs-gravity coupling is obtained by choosing the no-scale K\"ahler potential \cite{Ketov:2012jt}
\be \lb{noscale}
K(H,\bar{H})= -3 \ln\left[ X(H)+\bar{X}(\bar{H})\right] \quad {\rm with} \quad X(H)=\x H^2~, 
\ee
together with the Wess-Zumino-type superpotential $W$  that is a cubic polynomial in $H$.

The inflaton field $\f$ related to the real part of the leading component of the superfield $H$ has a physical pseudo-scalar superpartner $\c$ that also has the non-minimal coupling to gravity \cite{Ketov:2012jt},
\be \lb{ncsu}
S_{\rm non-minimal} = \fracmm{\x}{6}\int d^4x\; \sqrt{-g} (\f^2-\c^2)R~.
\ee
Therefore, the dynamics of $\c$ has to be suppressed during inflation, which requires extra tools as i.e., an extra superfield 
as the stabilizer and extra interactions just for the stabilization.

As is clear from the above arguments, the main problem is related to the complexification of inflaton in a chiral supermultiplet. However, inflaton can also belong to a {\it massive vector} supermultiplet $V$ that has a {\it single\/} physical scalar. The scalar potential of a vector multiplet is given by the $D$-term instead of the  $F$-term, while any desired values of the CMB observables ($n_s$ and $r$) are derivable from the inflaton potential proportional to the derivative squared of arbitrary real potential $J(V)$ \cite{Farakos:2013cqa,Ferrara:2013rsa}. The Lagrangian in curved superspace of supergravity is given by
\begin{equation}
\label{hsslag}
\mathcal{L}=\int d^2\theta 2\mathcal{E}\left\lbrace \frac{3}{8}(\overbar{\mathcal{D}}\overbar{\mathcal{D}}-8\mathcal{R})e^{-\frac{2}{3}J}+\frac{1}{4}W^\alpha W_\alpha
\right\rbrace +{\rm h.c.}~,
\end{equation}
and its bosonic part in Einstein frame (after Weyl rescalings) reads  \cite{Farakos:2013cqa,Ferrara:2013rsa}.
\begin{equation} \label{complag}
e^{-1}\mathcal{L}=\frac{1}{2}R -\frac{1}{4}F_{mn}F^{mn}-
\frac{1}{2}J''\partial_mC\partial^mC-\frac{1}{2}J''B_mB^m-\frac{g^2}{2}{J'}^2~,
\end{equation}
where $C=\left. V\right|$ is the real scalar inflaton field and $J=J(C)$.

The D-type scalar potential of the Starobinsky inflationary model is obtained by choosing the $J$ potential as
\begin{equation} \label{starj}
 J(C)=  \frac{3}{2} \left( C- \ln C\right)\quad {\rm and} \quad 
 C =  \exp\left( \sqrt{2/3} \phi\right)~.
 \end{equation}

\section{Equivalence of the Starobinsky and Higgs inflationary models in supergravity}

The Higgs inflation and its equivalence to the Starobinsky inflation in supergravity can be derived by using 
the minimal supergravity framework and the assignment of inflaton to a single massive vector multiplet, which is outlined in the end of the previous Section.

Let us consider the master function (potential) $J(V)$ as a function $\tilde{J}(He^{2V}\overbar{H})$, where we have introduced the Higgs chiral superfield $H$ and have chosen $g=1$ for simplicity. The argument of the function $\tilde{J}$ and the function itself are  invariant under the gauge transformations 
\begin{equation}
 \label{sgtr}
H\to e^{-iZ}H~,\quad \overbar{H}\to e^{i\overbar{Z}}\overbar{H}~,\quad V\to 
V+ \frac{i}{2}(Z-\bar{Z})~,
\end{equation}
whose gauge parameter Z itself is a chiral superfield. The original theory of the massive vector multiplet governed by the master function $J$  is recovered in the supersymmetric gauge $H=1$. 

We can now choose another (Wess-Zumino) supersymmetric gauge in which $V=V_1$, where $V_1$ describes the irreducible {\it massless} vector multiplet minimally coupled to the {\it dynamical} Higgs chiral multiplet $H$. The standard Higgs mechanism appears when choosing the canonical function $J=\frac{1}{2} He^{2V}\bar{H}$ that corresponds to a linear function $\tilde{J}$ \cite{Aldabergenov:2016dcu,Aldabergenov:2017bjt}. This phenomenon is known in the literature as the super-Higgs effect \cite{Wess:1992cp}. In our case, the supergravity theory in terms of the superfields $H$ and $V_1$
defines the Higgs inflation in supergravity that is equivalent to the Starobinsky inflation by construction, because both appear in the two different gauges of the {\it same} supergravity theory. 

The difference between this Higgs inflation in supergravity with inflaton belonging to the chiral (charged) superfield $H$ against the standard supergravity framework (Sec.~5) is due to the gauge invariance: now the scalar superpartner of inflaton is  a gauge (unphysical) degree of freedom.

In order to illustrate these conclusions by a simple argument, the best strategy is to use the same approximation as 
in Sec.~4, i.e., consider large scalar fields and ignore their kinetic terms. Then the relevant part of the supergravity action (\ref{hsslag}) {\it before} Weyl rescaling to Einstein frame reads ($M_{\rm Pl}=1$)~\footnote{See 
Ref.~\cite{Aldabergenov:2016dcu} for its derivation.} 
\be \lb{relp}
e^{-1} {\cal L} = \exp{\left(-\frac{2}{3}J\right)} \left(\frac{1}{2} R\right) - \frac{1}{2} g^2 \exp{\left(-\frac{4}{3}J\right)} (J')^2~~,
\ee
where by using Eq.~(\ref{starj}) we have 
\be \lb{Omega} 
e^{-\frac{2}{3}J} =Ce^{-C}\equiv \Omega >0~.
\ee
Therefore, Eq.~(\ref{relp}) can be rewritten to the form
\be \lb{relp2}
e^{-1} {\cal L} = \O \left(\frac{1}{2} R\right) - \frac{1}{2} \left(\frac{3}{2}g\right)^2 \O^2 \left(1-C^{-1}\right)^2~, 
\ee
where the field $\Omega$ is auxiliary, and the field $C=C(\O)$ is the special function of $\Omega$, which related to
 Lambert function. 
 
 Varying the Lagrangian ${\cal L}$  with respect to $\O$ yields
\be \lb{OR}  
\frac{1}{2}  R = \left(\frac{3}{2}g\right)^2 \Omega \left( 1 - \fracmm{2}{C(\Omega)}\right)\approx
\left(\frac{3}{2}g\right)^2 \Omega \left( 1 + \fracmm{2}{\ln\Omega}\right)~~,
\ee
where in the {\it large} field approximation, $C^{-1}\ll 1$ and $|1/\ln\O| \ll 1$, so that in the leading order we find
\be \lb{leada}
\frac{1}{2}  R \approx \left(\frac{3}{2}g\right)^2\O~.
\ee
Substituting it back into the Lagrangian yields 
\be \lb{lead1}
e^{-1} {\cal L} \approx \frac{1}{8} \left(\frac{3}{2}g\right)^{-2} R^2
\ee
as the leading term, i.e. the $R^2$ inflation.

After including the {\it next-to-leading term}  we find
\be \lb{next}
e^{-1} {\cal L} \approx  \frac{1}{8} \left(\frac{3}{2}g\right)^{-2} R^2
\left[ 1+ \fracmm{2}{\ln\left( \frac{2}{9}R/g^{2}\right)}\right]~.
\ee
This gives rise to the modified $R^2$ inflation in the gauge $H=1$. Similarly, when using the Wess-Zumino gauge $V=V_1$ with the charged Higgs (Stueckelberg) superfield $H$  and the function $\tilde{J}(\bar{H}e^{2gV_1}H)$, 
the $R^2$ inflation is reproduced  after introducing another function $\exp[-\frac{2}{3}\tilde{J}(\bar{H}H)]=\tilde{\O}$ as the auxiliary field and ignoring both the $H$-kinetic term and  the gauge field dependence in $V_1$ in the large field approximation.

We conclude this Section with a few comments.

In the supergravity model defined by Eqs.~(\ref{hsslag}) and (\ref{complag}), supersymmetry is restored {\it after} inflation in a Minkowski vacuum because the extreme points of the scalar potential obey the condition $J'=0$, so that the scalar potential itself also vanishes in the vacuum. It is possible to avoid it and get a de Sitter vacuum after inflation by introducing a hidden sector described by the Polonyi chiral superfield with a linear superpotential \cite{Aldabergenov:2016dcu}.

When adding any number of chiral (matter) superfields $\Phi_i$ described by a K\"ahler potential $K(\Phi_i,\overbar{\Phi}_i)$ and a superpotential $W(\Phi_i)$, the {\it F-type} scalar potential is also generated and is given by
\cite{Aldabergenov:2016dcu}
\begin{equation} \lb{extv}
V_F=e^{K+2J}\left[K^{ij^*}(W_i+K_iW)(\overbar{W}_{j^*}+K_{j^*}\overbar{W})+\left(2\fracmm{J^2_C}{J_{CC}}-3\right)|W|^2\right]~,
\end{equation}
where $K^{ij^*}$ is the inverse of K\"ahler metric $K_{ij^*}$, the indices $i$ and $j^*$ stand for the derivatives with respect to $\Phi_i$ and $\overbar{\Phi}_j$, respectively, and the subscripts $C$ denote the derivatives with respect to $C$.
 Equation (\ref{extv}) is the extension of Eq.~(\ref{eng}) in the presence of 
a vector multiplet self-interaction governed by the potential $J$. 

There is another $\eta$-problem related to Eq.~(\ref{extv}): even when all the fields $\Phi_i$ are stabilized during inflation driven by the D-term, the pre-factor $e^{2J}$ in the Starobinsky case of the $F$-term leads to an instability of inflation due to the appearance of the double exponential of the canonical inflaton field because of Eq.~(\ref{starj}). This problem can be  solved by using the alternative (field-dependent) Fayet-Iliopoulos term, as was demonstrated in Ref.~\cite{Aldabergenov:2017hvp}.

\section{Conclusion}

The Starobinsky and Higgs inflationary models in gravity are distinguished by their simplicity: they have only one relevant parameter (say, the inflaton mass) that is fixed by observations also. Hence, at present, both models do not have free parameters for describing inflation at all. Nevertheless, they remain viable by providing the best fit to all the observational data available. It also means that both inflationary models have the maximal predictive power leading to sharp predictions for the values of still unknown inflationary observables such as the tensor-to-scalar ratio. 

These valuable features are extendable to supergravity, as we demonstrated in this paper, when using the proper supergravity framework. 

Though low non-gaussianity of the CMB radiation favors single-field inflationary models, it does not exclude possible mixing  of inflaton with other scalars. However, this mixing should be suppressed, which is non-trivial in supergravity.
The Starobinsky-Higgs mixing is also possible \cite{Kaneda:2015jma,He:2018gyf,Gundhi:2018wyz,greece}, as well as small deviations from the basic $R^2$ inflation, caused by extra degrees of freedom present during inflation and quantum gravitational corrections.

\section*{Acknowledgements}

The author is supported in part by Tokyo Metropolitan University, the Competitiveness Enhancement Program of Tomsk Polytechnic University in Russia, and the World Premier International Research Center Initiative (WPI Initiative), MEXT, Japan.  The author is grateful to Alexei A. Starobinsky for suggesting the problem given by the title of this paper, and 
Andrei O. Barvinsky, Gia Dvali, Ioannis Gialamas and Christian Steinwachs for discussions and correspondence. The author is also grateful to the Mainz Institute for Theoretical Physics (MITP) of the DFG Cluster of Excellence PRISMA${}^+$ (Project ID 39083149), for its hospitality and its partial support during the completion of this work.  
\vglue.2in

\bibliographystyle{utphys} 

\begin{thebibliography}{10}

\bibitem{Starobinsky:1980te}
  A.~A.~Starobinsky,
  ``A New type of isotropic cosmological models without singularity,''
  Phys.\ Lett.\  {\bf B91} (1980) 99.

\bibitem{Bezrukov:2007ep}
F.~L.~Bezrukov  and M.~Shaposhnikov,
"The Standard Model Higgs boson as the inflaton",
  \href{http://dx.doi.org/10.1016/j.physletb.2007.11.072}{{Phys. Lett}
  {\bfseries B659} (2008) 703}
[arXiv:0710.3755 [hep-th]].


\bibitem{mf}  Y. Fujii and K-I. Maeda,
"The scalar-tensor theory of gravitation", Cambridge: Cambridge University Press, 2007, 260 p.


\bibitem{Aldabergenov:2018qhs} 
Y.~Aldabergenov, R.~Ishikawa, S.~V.~Ketov and S.~I.~Kruglov,
"Beyond Starobinsky inflation", 
\href{http://dx.doi.org/10.1103/PhysRevD.98.083511}{{Phys. Rev.} {\bfseries D98} (2018) 083511}
[arXiv:1807.08394  [hep-th]].

\bibitem{Ketov:2012yz}
S.~V. Ketov, ``{Supergravity and early Universe: the meeting point of cosmology
  and high-energy physics},''
  \href{http://dx.doi.org/10.1142/S0217751X13300214}{{Int. J. Mod. Phys.}
  {\bfseries A28} (2013) 1330021}
[arXiv:1201.2239 [hep-th]].


\bibitem{starp} A.~A.~Starobinsky, "Cosmological models of primordial and present dark energy in $f(R)$ gravity",
Lectures at the 49th Winter School, "Cosmology and non-equilibrium statistical mechanics", Ladek-Zdroj, Poland, 13--15 February 2013 (unpublished).

\bibitem{Ketov:2014kya}
S.~V.~Ketov and N.~Watanabe,
"On the Higgs-like Quintessence for Dark Energy",
\href{http://dx.doi.org/10.1142/S021773231450117X}{{Mod. Phys. Lett.} {\bfseries A29} (2014) 1450117}
[arXiv:1401.7756  [hep-th]].

\bibitem{Ketov:2014jta} 
S.~V.~Ketov and N.~Watanabe, 
"The $f(R)$ gravity function of Linde quintessence",
\href{http://dx.doi.org/10.1016/j.physletb.2014.12.047}{{Phys. Lett.} {\bfseries B741} (2015) 242}
[arXiv:1410.3557  [hep-th]].

\bibitem{Ellis:2015pla} 
J.~Ellis, M.~A.~G. Garcia, D.~V. Nanopoulos and K.~A. Olive,
"Calculations of inflaton decays and reheating: with applications to no-scale inflation models",
 JCAP {\bf 1507} (2015) 050                     
[arXiv:1505.06986 [hep-ph]].                       

\bibitem{Planck18} Y. Akrami et al. [Planck Collaboration], 
"Planck 2018 results. X. Constraints on inflation", [arXiv:1807.06211 [astro-ph.CO]].

   \bibitem{Mukhanov:1981xt}
V.~F. Mukhanov and G.~V. Chibisov, 
"Quantum fluctuations and a nonsingular universe",
JETP Lett. {\bf 33} (1981) 532 [Pisma Zh. Eksp. Teor. Fiz. {\bf 33} (1981) 549].

\bibitem{Kaneda:2010ut}
S.~Kaneda, S.V. Ketov and N. Watanabe, 
"Fourth-order gravity as the inflationary model revisited",
Mod. Phys. Lett. {\bf A25} (2010) 2753
[arXiv:1001.5118 [hep-th]].                

\bibitem{Hertzberg:2010dc}
M.~P.~Hertzberg,
 "On inflation with non-minimal coupling",
  \href{http://dx.doi.org/10.1007/JHEP11(2010)023}{{JHEP}
  {\bfseries 1011} (2010) 023}
[arXiv:1002.2995 [hep-ph]].
 
\bibitem{Ruf:2017xon}
M.~S.~Ruf and C.~F.~Steinwachs,
"Quantum equivalence of $f(R)$ gravity and scalar-tensor
                        theories",
    \href{http://dx.doi.org/10.1103/PhysRevD.97.044050}{{Phys. Rev.}
  {\bfseries D97} (2018) 044050}
[arXiv:1711.07486 [gr-qc]].                      
                        

\bibitem{Burgess:2009ea}
C.~P.~Burgess, H.~M.~Lee and M.~Trott,
"Power-counting and the validity of the classical
                        approximation during inflation",
          \href{http://dx.doi.org/10.1088/1126-6708/2009/09/103}{{JHEP}
  {\bfseries 0909} (2009) 103}
[arXiv:0902.4465 [hep-ph]].                
                        
    
\bibitem{Ketov:2012jt}
S.~V.~Ketov  and A.~A.~Starobinsky,
"Inflation and non-minimal scalar-curvature coupling in
                        gravity and supergravity",
   \href{http://dx.doi.org/10.1088/1475-7516/2012/08/022}{{JCAP}
  {\bfseries 1208} (2012) 022}
[arXiv:1203.0805 [hep-th]].

\bibitem{Kehagias:2013mya}              
   A.~Kehagias, A.~M.~Dizgah and A.~Riotto,
      "Remarks on the Starobinsky model of inflation and its
                        descendants",
    \href{http://dx.doi.org/10.1103/PhysRevD.89.043527}{{Phys. Rev.}
  {\bfseries D89} (2014) 043527}
[arXiv:1312.1155 [hep-th]].
                      
                                             
\bibitem{Ketov:1995yd}
S.~V.~Ketov, "Conformal Field Theory", Singapore: World Scientific (1995), 486 p.

\bibitem{Yamaguchi:2011kg}
M.~Yamaguchi, ``{Supergravity based inflation models: a review},''
  \href{http://dx.doi.org/10.1088/0264-9381/28/10/103001}{{Class. Quant.
  Grav.} {\bfseries 28} (2011) 103001}
[arXiv:1101.2488  [astro-ph.CO]].


\bibitem{Ketov:2014qha}
S.~V.~Ketov and T.~Terada,
 "Inflation in supergravity with a single chiral
                        superfield",                  
     \href{http://dx.doi.org/10.1016/j.physletb.2014.07.036}{{Phys. Lett.}
  {\bfseries B736} (2014) 272}
[arXiv:1406.0252 [hep-th]].
                     
        
        
        
\bibitem{Ketov:2014hya} 
S.~V.~Ketov and T.~Terada,
"Generic scalar potentials for inflation in supergravity
                        with a single chiral superfield",
     \href{http://dx.doi.org/10.1007/JHEP12(2014)062}{{JHEP}
  {\bfseries 1412} (2014) 062}
[arXiv:1408.6524 [hep-th]].                    
                        
 

\bibitem{Ketov:2010qz}
S.~V.~Ketov  and A.~A.~Starobinsky,
"Embedding $(R+R^2)$ inflation into supergravity",
   \href{http://dx.doi.org/10.1103/PhysRevD.83.063512}{{Phys. Rev.}
  {\bfseries D83} (2011) 063512}
[arXiv:1011.0240 [hep-th]].                 

                          
     

\bibitem{Ketov:2013dfa}
S.~V.~Ketov and T.~Terada,
"Old-minimal supergravity models of inflation",
  \href{http://dx.doi.org/10.1007/JHEP12(2013)040}{{JHEP}
  {\bfseries 1312} (2013)  040}
[arXiv:1309.7494 [hep-th]].

\bibitem{Wess:1992cp}
 J, Wess and J. Bagger,  "Supersymmetry and Supergravity",
Princeton: Princeton  University Press, 1992, 2nd Edition, 260 p.

\bibitem{Cecotti:1987sa}
S.~Cecotti, "Higher-derivative supergravity is equivalent to standard
supergravity coupled to matter,
  \href{http://dx.doi.org/10.1016/0370-2693(87)90844-6}{{Phys. Lett.}
  {\bfseries B190} (1987)  86}.


\bibitem{Gates:2009hu}
 S.~J.~Gates,~Jr., and S.~V.~Ketov,
  "Superstring-inspired supergravity as the universal
                        source of inflation and quintessence",
                        \href{http://dx.doi.org/10.1016/j.physletb.2009.03.005}{{Phys. Lett.}
  {\bfseries B674} (2009)  59}
  [arXiv:0901.2467 [hep-th]].

                     
\bibitem{Farakos:2013cqa}
   F.~Farakos, A.~Kehagias and A.~Riotto,
   "On the Starobinsky model of inflation from
                        supergravity",
    \href{http://dx.doi.org/10.1016/j.nuclphysb.2013.08.005}{{Nucl. Phys.}
  {\bfseries B876} (2013)  187}
[arXiv:1307.1137 [hep-th]].

                       
\bibitem{Ferrara:2013rsa}   
S.~Ferrara, R.~Kallosh, A.~Linde and M.~Porrati,
"Minimal supergravity models of inflation",
  \href{http://dx.doi.org/10.1103/PhysRevD.88.085038}{{Phys. Rev.}
  {\bfseries D88} (2013)  085038}
[arXiv:1307.7696 [hep-th]].

                  
\bibitem{Aldabergenov:2016dcu}
  Y.~Aldabergenov and S.~V.~Ketov,
  ``SUSY breaking after inflation in supergravity with inflaton in a massive vector supermultiplet,''
     \href{http://dx.doi.org/10.1016/j.physletb.2016.08.016}{{Phys. Lett.}
  {\bfseries B761} (2016)  115}
[arXiv:1607.05366 [hep-th]].
 

\bibitem{Aldabergenov:2017bjt} 
  Y.~Aldabergenov and S.~V.~Ketov,
  ``Higgs mechanism and cosmological constant in $N=1$
                        supergravity with inflaton in a vector multiplet,''
       \href{http://dx.doi.org/10.10.1140/epjc/s10052-017-4807-8}{{Eur. Phys. J.}
  {\bfseries C77} (2017)  233}
[arXiv:1701.08240 [hep-th]].
                      
                     
                        
                       
\bibitem{Aldabergenov:2017hvp}
  Y.~Aldabergenov and S.~V.~Ketov,
  "Removing instability of inflation in
                        Polonyi-Starobinsky supergravity by adding FI term",
    \href{http://dx.doi.org/10.1142/S0217732318500323}{{Mod. Phys. Lett.}
  {\bfseries A91} (2018)  1850032}
[arXiv:1711.06789 [hep-th]].
                      
                        


\bibitem{Kaneda:2015jma}
S.~Kaneda  and S.~V.~Ketov,
 "Starobinsky-like two-field inflation",
  \href{http://dx.doi.org/10.140/epjc/s10052-016-3888-0}{{Eur. Phys. J.}
  {\bfseries C76} (2016)  26}
[arXiv:1510.03524 [hep-th]].

  
\bibitem{He:2018gyf}
M.~He, A.~A.~Starobinsky and J.~Yokoyama,
"Inflation in the mixed Higgs-$R^2$ model",
  \href{http://dx.doi.org/10.1088/1475-7516/2018/05/064}{{JCAP}
  {\bfseries 1805} (2018)  064}
[arXiv:1804.00409 [hep-th]].

\bibitem{Gundhi:2018wyz}
 A.~Gundhi and C.~F.~Steinwachs,
"Scalaron-Higgs inflation",
[arXiv:1810.10546 [hep-th]].

\bibitem{greece} D.~D.~Canko, I.~D.~Gialamas and G.~P.~Kodaxis, 
"A simple $F({\cal R},\phi)$ deformation of Starobinsky inflationary model",
[arXiv:1901.06296 [hep-th]].
 


\end{thebibliography}

\providecommand{\href}[2]{#2}\begingroup\raggedright
\endgroup

\end{document}